\documentclass[10pt, conference, compsocconf]{IEEEtran}
\pdfoutput=1 %for posting to arXiv
\usepackage{blindtext, graphicx}
\usepackage{epsfig}
\usepackage{syntonly}
\usepackage{rotating}
\usepackage{amsmath}
\usepackage{setspace}
\usepackage{verbatim} 
\usepackage{amssymb}
\usepackage{amsmath}
\usepackage{graphicx}
\usepackage{multirow}
\usepackage{colortbl}
\usepackage{enumerate}
\usepackage{color}
\usepackage{wrapfig}
\usepackage{caption}
\usepackage{float}

\floatstyle{plaintop}
\restylefloat{table}
\usepackage{algorithm2e}
\usepackage{comment}
\usepackage{ifthen}
\usepackage[dotinlabels]{titletoc}
\usepackage{colortbl}
\usepackage{array}
\usepackage{alltt}
\usepackage{bm}

\usepackage[normalem]{ulem}

\pdfoutput=1 %for posting to arXiv
\usepackage{blindtext, graphicx}
\usepackage{epsfig}
\usepackage{syntonly}
\usepackage{rotating}
\usepackage{amsmath}
\usepackage{setspace}
\usepackage{verbatim} 
\usepackage{amssymb}
\usepackage{amsmath}
\usepackage{graphicx}
\usepackage{multirow}
\usepackage{colortbl}
\usepackage{enumerate}
\usepackage{color}
\usepackage{wrapfig}
\usepackage{algorithm2e}
\usepackage{hyperref}
\usepackage{cleveref}

\usepackage{comment}
\usepackage{ifthen}
\usepackage[dotinlabels]{titletoc}
\usepackage{colortbl}
\usepackage{array}

\usepackage{alltt}
\usepackage{bm}
\usepackage{longtable,booktabs}
\usepackage{url}
\usepackage{courier}
\usepackage{cite}

\newcommand{\authemail}[1]{\normalfont{\texttt{#1}}}

\newcommand{\Affil}[1]{\fontsize{11}{11}\itshape{#1}}
\usepackage{tabularx}
\usepackage{footnote}

\begin{document}
\title{Intra-model Variability in COVID-19 Classification Using Chest X-ray Images}

\author{
\IEEEauthorblockN{
	\large{Brian D Goodwin\IEEEauthorrefmark{1}}
	Corey Jaskolski\IEEEauthorrefmark{1}
	Can Zhong\IEEEauthorrefmark{1} 
	Herick Asmani\IEEEauthorrefmark{1}
}
\IEEEauthorblockA{\Affil{\IEEEauthorrefmark{1}Synthetaic}
%\IEEEauthorblockA{\Affil{\IEEEauthorrefmark{1}SynthetAIc}, 
%\IEEEauthorrefmark{2}NVIDIA
}
\authemail{\{brian,corey,can,herick\}@synthetaic.com}
}
\maketitle

\begin{abstract}

X-ray and computed tomography (CT) scanning technologies for COVID-19 screening have gained significant traction in AI research since the start of the coronavirus pandemic. Despite these continuous advancements for COVID-19 screening, many concerns remain about model reliability when used in a clinical setting. Much has been published, but with limited transparency in expected model performance. We set out to address this limitation through a set of experiments to quantify baseline performance metrics and variability for COVID-19 detection in chest x-ray for 12 common deep learning architectures. Specifically, we adopted an experimental paradigm controlling for train-validation-test split and model architecture where the source of prediction variability originates from model weight initialization, random data augmentation transformations, and batch shuffling. Each model architecture was trained 5 separate times on identical train-validation-test splits of a publicly available x-ray image dataset provided by Cohen et al. (2020). Results indicate that even within model architectures, model behavior varies in a meaningful way between trained models. Best performing models achieve a false negative rate of 3 out of 20 for detecting COVID-19 in a hold-out set. While these results show promise in using AI for COVID-19 screening, they further support the urgent need for diverse medical imaging datasets for model training in a way that yields consistent prediction outcomes. It is our hope that these modeling results accelerate work in building a more robust dataset and a viable screening tool for COVID-19.
\end{abstract}

\begin{IEEEkeywords}
 Deep Learning; COVID-19; X-ray; Screening;

\end{IEEEkeywords}

\IEEEpeerreviewmaketitle

\section{Introduction}
The spread of the novel coronavirus, which causes COVID-19, has caught most of the world off-guard resulting in severely limited testing capabilities. For example, as of April 15, 2020 almost 3 months since the first case in the US, only about 3.3 million tests have been administered \cite{trackingproject}, which equates to approximately 1\% of the US population. Reverse transcription-polymerase chain reaction (RT-PCR) is an assay commonly used to test for COVID-19, but is available in extremely limited capacity \cite{ai2020correlation,shi2020review}. In an effort to offer a minimally invasive, low-cost COVID-19 screen via x-ray imaging, AI engineers and data scientists have begun to collect datasets \cite{cohencovid192020} and utilize computer vision and deep learning algorithms \cite{wangcovidnet2020}. All these efforts seek to leverage an available medical imaging modality for both diagnosis and, in the future, predicting case outcome. Clinical observations have largely propelled AI research in computer vision for screening COVID-19, and these reports cite differentiable lung abnormalities of COVID-19 patients from chest CT~\cite{yan2020covid19}, x-ray~\cite{huang_clinical_2020, ng2020imaging}, and even ultrasound \cite{born2020pocovidnet}. Current research also shows that COVID-19 is correlated with specific biomarkers in x-ray~\cite{apostolopoulos2020extracting}.
 
Though these recent efforts are valuable in that they will lay the foundation for future work in this area, there are significant flaws in the methodology as well as in the behavior of the resultant models. Much of the initial work on COVID-19 prediction from chest x-ray used a training set that included a little over 100 images with 10 test images (that were, in fact, identical to the validation set). Though such small test data sets do not allow for declaring sweeping diagnostic value statements, unfortunately the popular media articles effectively hype the value of these models with hopeful titles like "Coronavirus Neural Network can help spot COVID-19 in Chest X-rays" \cite{canhelp}, "How AI Is Helping in the Fight Against COVID-19" \cite{aihelps}, and "A.I. could help spot telltale signs of coronavirus in lung X-rays" \cite{aicouldspot}.

Network weights from these publications are not publicly available for these published models. We have responded to this shortcoming by providing pre-trained weights for many of the most common deep learning architectures for computer vision, and we have made the code for pre-training freely available. To our knowledge, this repository of pre-trained model weights is the first of its kind in response to the current crisis and the first to report prediction results across multiple architectures on a test set that is held out from the validation and training sets.
 
Our goal is to facilitate advancement of screening technology for COVID-19 and highlight the need for larger, more diverse datasets. The urgency for a clinical methodology to provide COVID-19 screens cannot be understated \cite{kanne2020essentials}. Our hope is twofold: 1) that the community advances computer vision for COVID-19 detection via x-ray before recommending use in a clinical setting and 2) that pre-trained model weights will help accelerate ongoing development in AI to augment the decision-making process for clinicians during a time where healthcare workers are under a severe amount of stress.

\section{Methods}
\label{sec:methods}

We carried out a series of experiments to quantify baseline machine-learning performance in detecting COVID-19 from chest x-ray images using a series of common, openly available neural network architectures. Computational benchmarking was outside of our experimental approach since it has been extensively studied \cite{bianco_benchmark_2018}. In this study, we focused on quantifying the expected variability in prediction outcomes and sought to quantify the reliability of predictions with respect to the chest x-ray data that is currently available to the public and contains COVID-19 positive scans. All computational experiments were executed on Lambda Blade (lambdalabs.com) hardware (8x RTX 8000 + NVLink GPUs with 48GB GDDR6 RAM, 2x Intel Xeon Gold 5218 processor with 512 GB RAM) using the Pytorch framework \cite{NEURIPS2019_9015}.

\subsection{Experimental Design}

An identical train-validation-test split (TVTS) was employed for all experiments, and each model architecture was trained 5 separate times (creating 5 separate models, each). We controlled for TVTS and model architecture while allowing randomness in weight initialization and data augmentation during batch training for each experiment. We designed our approach with the aim to elucidate expected variability in model behavior for COVID-19 detection in chest x-ray.

\subsection{Data}

\begin{figure*}[htb]
\begin{center}
%[height=1in,width=1in,angle=-90]
\includegraphics{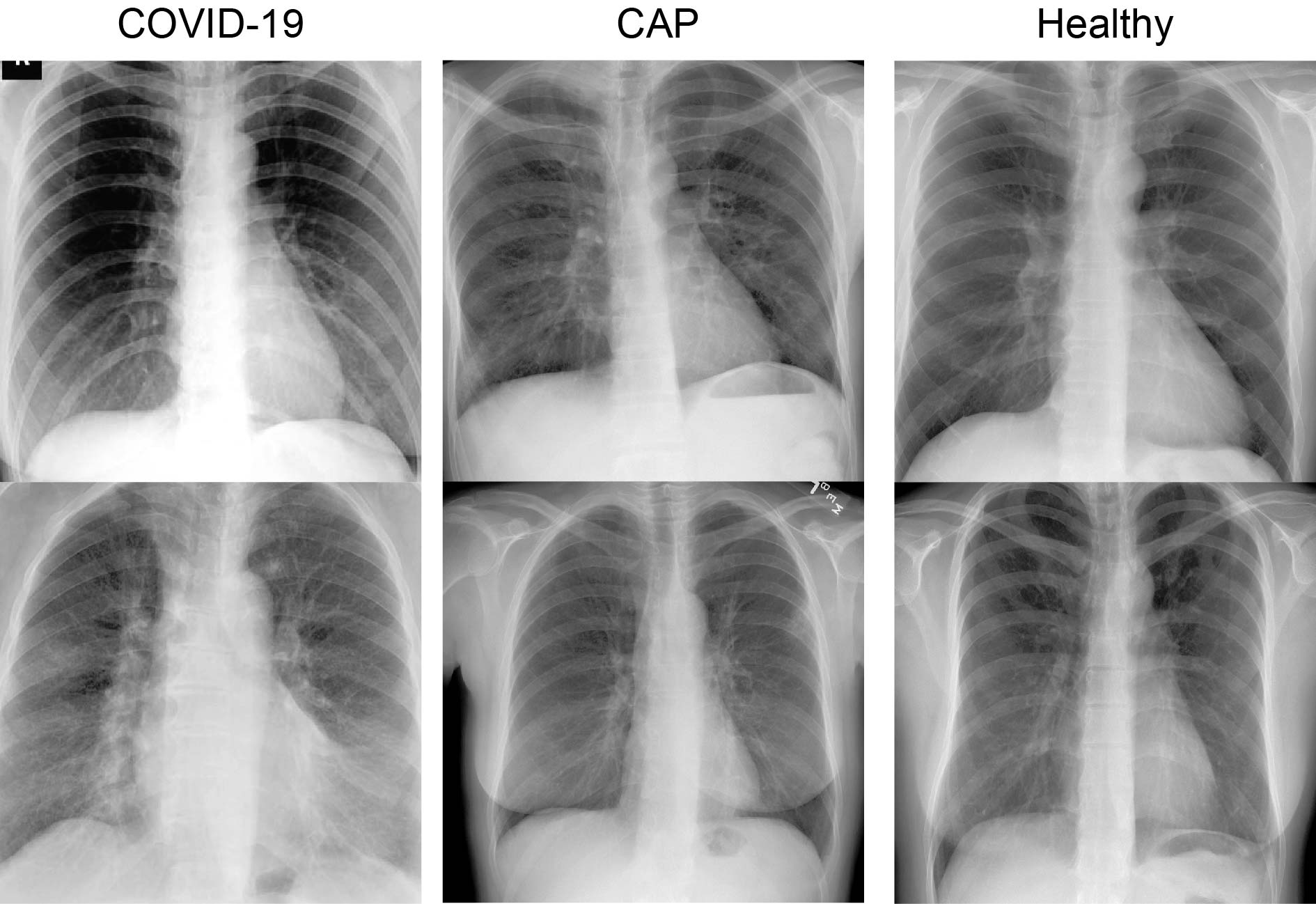}
\caption{Two example images from each class in the dataset.}
\label{fig:exampleimages}
\end{center}
\end{figure*}

Since COVID-19 datasets are becoming more abundant, existing sets are constantly growing and evolving, and it has become important to cite the data source and the day it was acquired. Specifically, this study uses data from the Cohen et al. \cite{cohencovid192020} chest x-ray dataset that was acquired on 2020-04-17. This dataset contains three classes: 1) healthy, 2) community acquired pneumonia (CAP), and 3) COVID-19 (examples in Figure \ref{fig:exampleimages}). We employed a modified 80-10-10 TVTS (Table \ref{tab:trainvaltestsplit}) training paradigm. It was modified to maximize the number of COVID-19 training samples and double the size of the test hold out set used in previous work \cite{wangcovidnet2020}. The dataset version at the time of this study was large enough to accommodate the addition of a COVID-19 validation set and test hold out set 2x larger than previous work (from 10 hold out images to 20) \cite{wangcovidnet2020}.

\begin{table}[ht]
    \centering
    \begin{tabular}{rrrr}
        \hline
        & COVID-19 & CAP & Normal \\ 
        \hline
        Train & 175 & 4836 & 7081 \\ 
        Validation & 20 & 605 & 885 \\ 
        Test & 20 & 604 & 885 \\ 
        \hline
    \end{tabular}
    \caption{Number of x-ray images in each split for training, validation, and hold-out testing.} 
    \label{tab:trainvaltestsplit}
\end{table}

\subsection{Architectures}

We elected to generate baseline results for the following commonly used architectures: Resnet-18, -50, -101, -152 \cite{he2016deep}, WideResnet-50, -101 \cite{corrZagoruykoK16}, ResNeXt-50, -101 \cite{xie2017aggregated}, MobileNet-v1 \cite{howard2017mobilenets}, Densenet-121, -169, -201 \cite{huang2017densely}. 

ADAM optimization was used during training on only the last fully connected layer in each network using a batch size of 128 resulting in a mean compute time of 156.7+/-50.7 sec/epoch across all architectures. Since all models have been pre-trained on ImageNet, we elected to freeze the convolutional layers to retain the higher-level learned features \cite{deng2009imagenet}; i.e., all weights were frozen but for the final layer in each network. Models were trained on chest x-ray images of size 3 x 512 x 512 px. All images were assumed to be RGB channels despite their inherent grayscale property. Therefore, no manipulations were made to the networks to uniquely accommodate single channel x-ray imaging data.

\begin{figure}[htb]
\begin{center}
%[height=1in,width=1in,angle=-90]
\includegraphics{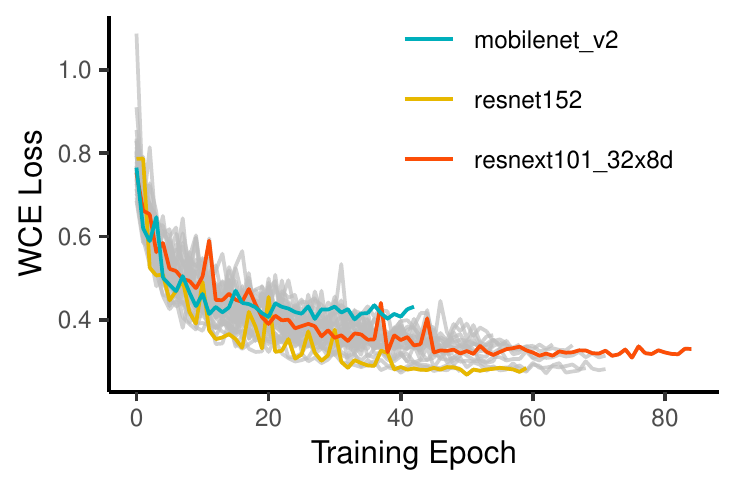}
\caption{Validation losses from trained epochs for all models with a select few shown by their respective line color.}
\label{fig:validation_loss}
\end{center}
\end{figure}

All models were trained for 100 epochs with stopping criteria, and weights from lowest validation loss were saved out; COVID-19 recall was \emph{not} considered during training. All experiments were carried out using weighted cross entropy loss (wCEL) where the contribution to the loss from a given class is weighted based on its representative proportion in the total dataset. 
\begin{equation*}
\resizebox{\columnwidth}{!}{$loss(x,class)=weight[class]\left ( -x[class] + \textup{log} \left ( \sum_{j} \textup{exp}(x[j]) \right ) \right )$}
\end{equation*}
Our decision to use wCEL (as opposed to CEL) is based on the objective to achieve a high recall in the underrepresented class (the COVID-19 class) since the AI task is straightforward: detect COVID-19. Stopping criteria was based on plateau of the validation wCEL. Performance metrics were then calculated using only the test holdout set. Our first iteration of testing (not reported in this paper) used CEL, and performance metrics were found to have improved dramatically when models were trained using wCEL (as reported in this paper). Validation losses followed a common trend across models during training with a distribution illustrated in Figure \ref{fig:validation_loss}).

\subsection{Performance Indices}

Each training experiment was carried out on an identical TVTS of the dataset. For each architecture, 5 training runs were carried out to gather a small distribution of models (and prediction outcomes). The class prediction (and therefore COVID-19 detection) was based solely on the maximum value output from the softmax layer (3 total classes). Sources of variability across all experiments included image augmentation transformations that are based on random draws from a binomial probability distribution, shuffling for batch allocation (i.e., each batch ID does not contain identical images across all experiments), and random weight initialization (last layer only). 

Datasets were prepared in a manner consistent with Wang et al. \cite{wangcovidnet2020} (see github.com/lindawangg/COVID-Net), and a data augmentation protocol was implemented. Given the consistent format of a chest x-ray, only modest translational and rotational augmentations were applied with brightness jitter and a possibility of horizontal flip (Table \ref{tab:dataaugmentation}).

\begin{table}[ht]
\centering
\begin{tabular}{lll}
  \hline
Augmentation & Range & Probability \\ 
  \hline
Horizontal Flip & & 0.5 \\ 
  Brightness Jitter & [0.9, 1.1]*value & 0.5 \\ 
  Random Rotation & [-10, 10] degrees & 0.5 \\ 
  Random Translation & [-0.1, 0.1]*size & 0.5 \\ 
   \hline
\end{tabular}
\caption{Data augmentation composition.} 
\label{tab:dataaugmentation}
\end{table}

\subsection{Statistical Analysis}
Multiple measures of accuracy and uncertainty were calculated to quantify baseline performance expectations for each network. We report statistical tests, model performance characteristics, and common accuracy metrics with the aim to quantify the expected performance and the variability of model performance given the size of current COVID-19 chest x-ray datasets. Specifically, we report Type I (false positive; FP) and Type II (false negative; FN) error frequencies in the test set as well as model consistency via McNemar’s chi-squared test (with a “continuity correction”), which is a common way to test for similarity between models \cite{dietterich1998,everitt1977}. This statistical test was carried out using the method described in Dietterich \cite{dietterich1998} for a binary classification task (COVID-19 v. non-COVID-19). All statistical analyses were carried out using R \cite{rcitation}, and figures were built using ggplot2 \cite{ggplot2}.

\section{Results}
\label{sec:results}

\begin{figure}[htb]
\begin{center}
%[height=1in,width=1in,angle=-90]
\includegraphics{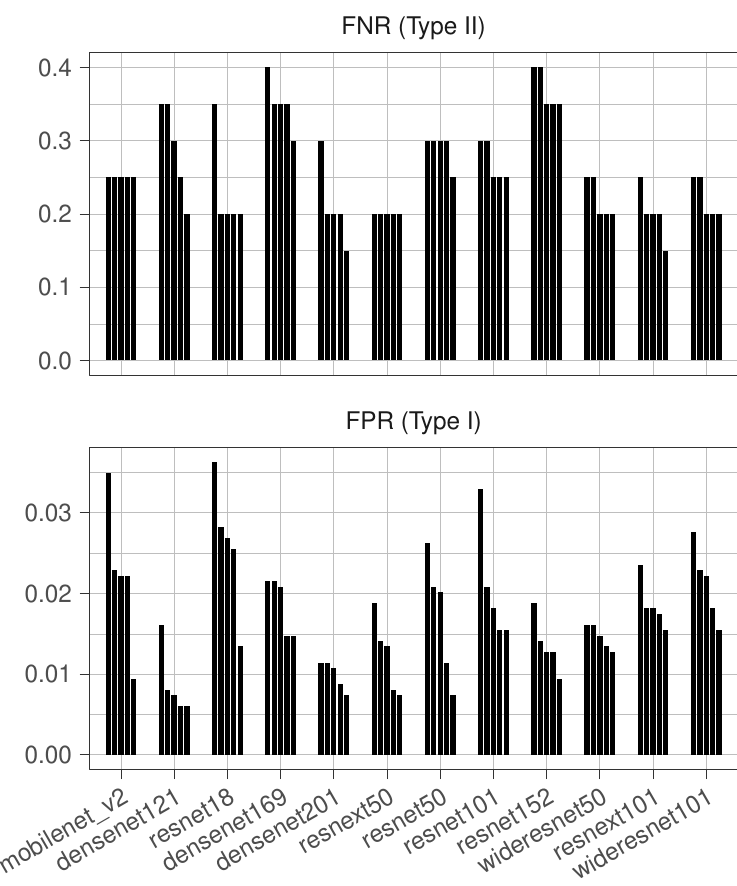}
\caption{Type I (FPR) and II (FNR) error rates for all tested architectures across all experiments. Each bar represents a single experiment (5 for each architecture).}
\label{fig:model_variability_fpr_fnr}
\end{center}
\end{figure}

\begin{figure*}[htb]
\begin{center}
%[height=1in,width=1in,angle=-90]
\includegraphics{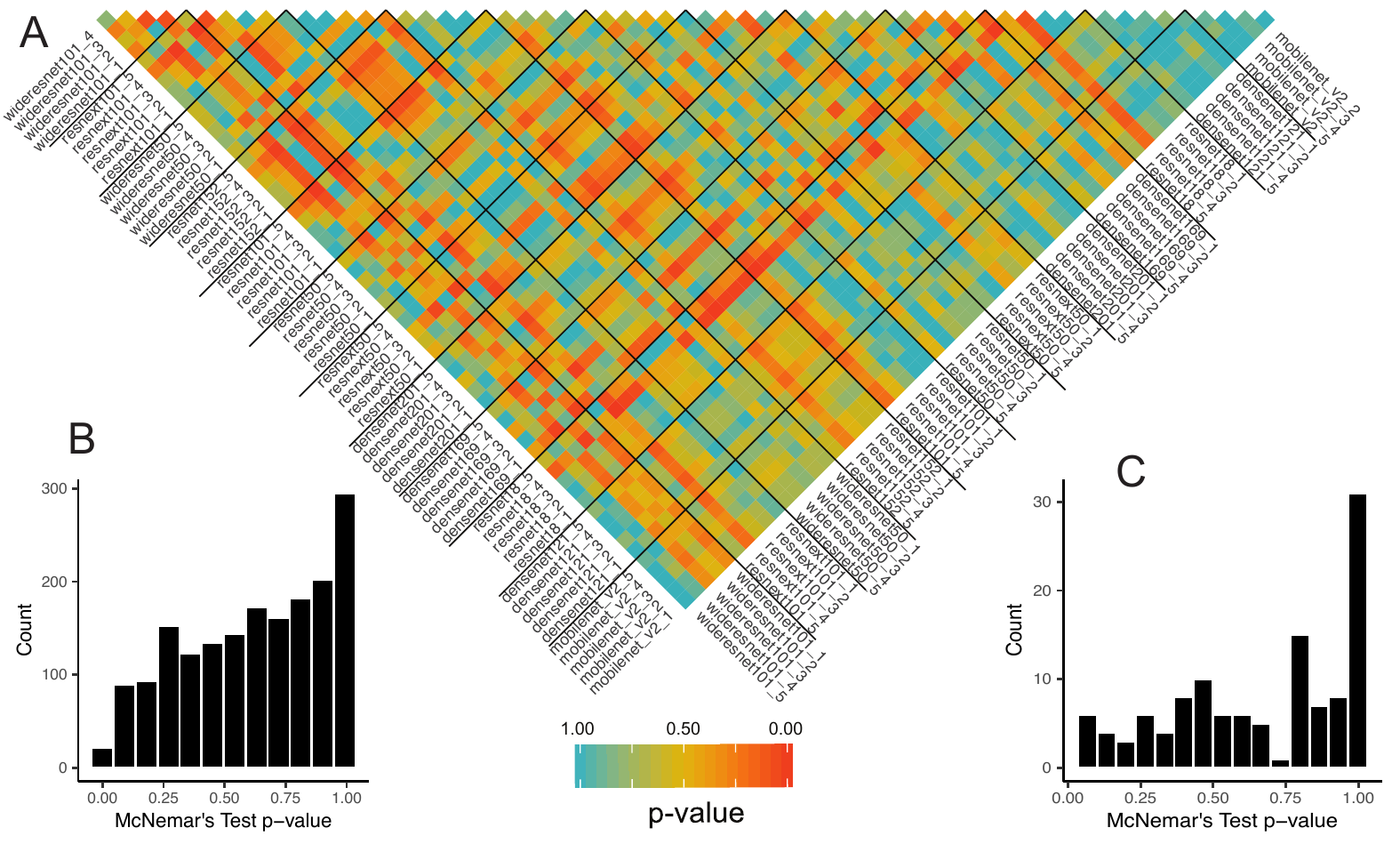}
\caption{A representation of p-values from all two-sample McNemar’s chi-squared tests is shown in 3 plots. The classification task was COVID-19 v. non-COVID-19 (binary). Tests were carried out between each model (5 for each architecture) and results are represented via histograms (B, C) and similarity matrix (A). Histogram (B) represents the count of all p-values over their respective values, and histogram (C) represents the count of p-values from tests where the two models in question have identical architectures. Colors in each tile within the lower triangle of the similarity matrix in (A) represent the p-value from the paired test. P-values near zero indicate that model behavior is \emph{different} while those near 1 indicate \emph{similar} behavior.}
\label{fig:mcnemar_v2}
\end{center}
\end{figure*}

Among all tested models, Densenet-169 was found to have the highest false negative rate (FNR), Densenet-121 had the highest FNR variance, and Resnet-18 had the highest mean false positive rate (FPR) (Figure \ref{fig:model_variability_fpr_fnr}). All plot labels for model architectures are organized by number of tunable parameters increasing from left to right (order reversed in Figure \ref{fig:mcnemar_v2}B). The lowest FNR (0.15) was achieved by Densenet-201 and ResneXt-101. Overall, Type I and II error rates varied across models and varied modestly within model architecture (Figure \ref{fig:model_variability_fpr_fnr}).

\begin{figure*}[htb]
\begin{center}
%[height=1in,width=1in,angle=-90]
\includegraphics{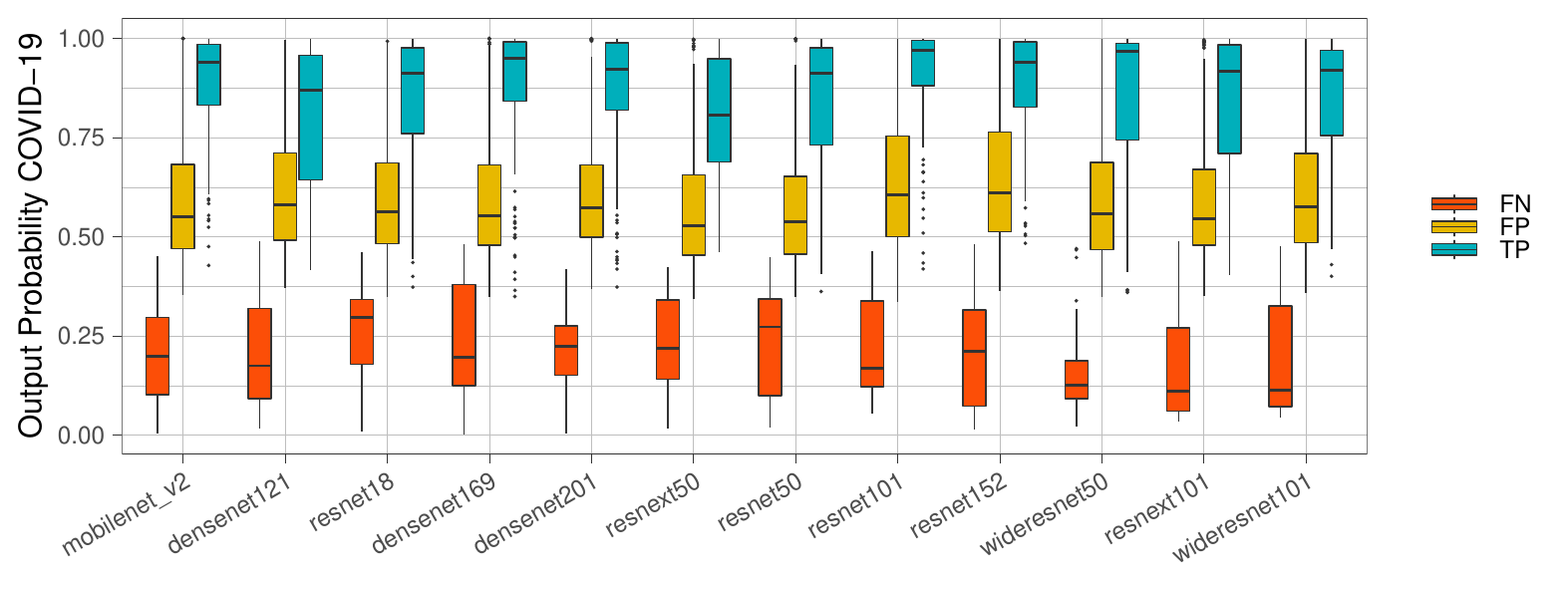}
\caption{Box-plot of output probability distributions from the softmax layer for true positive (TP), false positive (FP), and false negative (FN) cases.}
\label{fig:model_variability_covid_p_box}
\end{center}
\end{figure*}

Networks had consistently lower softmax output probabilities for COVID-19 in the event of Type I error (FP) while TP probability distributions consistently extended well into those from FP outputs (Figure \ref{fig:model_variability_covid_p_box}). No significant difference across architectures was found between softmax probabilities in the event of a more serious Type II error (FN; failing to correctly detect COVID-19). For the binary COVID-19 detection task, mean softmax outputs were 0.601+/-0.161 and 0.213+/-0.135 for FP and FN, respectively. Interestingly, 16.5\% of all TP predictions had softmax probability outputs below 0.667.

Prediction behavior was often inconsistent between models and those with identical architectures to a significant degree based on McNemar’s test (Figure \ref{fig:mcnemar_v2}). P-values of the McNemar’s chi-squared test statistic were used to estimate the prediction behavior consistency (in the test set only) between all models for the binary classification task of detecting COVID-19 versus non-COVID-19. With the null hypothesis, a low p-value suggests that the two models in question have inconsistent prediction behaviors. A large portion of model comparisons show low p-values and therefore high prediction inconsistency (Figure \ref{fig:mcnemar_v2}B). Similarly, low p-values were common when comparing models having identical architectures (Figure \ref{fig:mcnemar_v2}C). If prediction behavior were largely consistent between models, distributions shown in Figures \ref{fig:mcnemar_v2}B and C would have large spikes near 1.0, instead the distribution of p-values is more uniform than expected. Given that our experimental design controlled for the TVTS and model architecture across all experiments, it is expected that a high quality dataset would produce a distribution of p-values indicating similar behavior \emph{between models with identical architectures} (Figure \ref{fig:mcnemar_v2}C). McNamer’s test suggests that several architectures have reliably \emph{inconsistent} behavior. The most apparent differences are between Mobilenet-v2 v. Wideresnet-101, Resnet-152 v. Wideresnet-101, and Resnet-152 v. Wideresent-50 (Figure \ref{fig:mcnemar_v2}A). Mobilenet-v2 was found to have the least inter- and intra-model differences, which could be explained by its relatively small number of tunable parameters making a more "general" model fit. Predictions from Densenet-121 models had the most consistency, on average, with all other trained models including those with identical architecture. Models sharing the Wideresnet-101 architecture had the most intra-model differences followed by ResneXt-101. Comparisons between Mobilenet-v2, densenet-121, and Resnet-18 indicate that these architectures had the most similar prediction behavior (Figure \ref{fig:mcnemar_v2}B), and their false negative rate (FNR) values fall in the middle of the pack. Conversely, models with deeper architectures were responsible for the highest FNR values (Table \ref{tab:accmetrics}). 

\begin{table}[ht]
	\captionsetup{font=small}
	\centering
	\resizebox{\columnwidth}{!}{%
	\begin{tabular}{lrrrrrr}
	 \hline
	Architecture & TPR & FPR & FNR & PPV & F1 & ACC \\ 
	  \hline
	mobilenet\_v2 & 0.75 & 0.022 & 0.25 & 0.334 & 0.455 & 0.869 \\ 
	  densenet121 & 0.71 & \textbf{0.009} & 0.29 & \textbf{0.539} & 0.606 & 0.878 \\ 
	  resnet18 & 0.77 & 0.026 & 0.23 & 0.296 & 0.422 & 0.877 \\ 
	  densenet169 & 0.65 & 0.019 & 0.35 & 0.323 & 0.430 & 0.871 \\ 
	  densenet201 & 0.79 & 0.010 & 0.21 & 0.519 & \textbf{0.625} & \textbf{0.884} \\ 
	  resnext50 & \textbf{0.80} & 0.012 & \textbf{0.20} & 0.481 & 0.596 & 0.868 \\ 
	  resnet50 & 0.71 & 0.017 & 0.29 & 0.384 & 0.489 & 0.871 \\ 
	  resnet101 & 0.73 & 0.021 & 0.27 & 0.335 & 0.456 & 0.879 \\ 
	  resnet152 & 0.63 & 0.014 & 0.37 & 0.391 & 0.481 & 0.878 \\ 
	  wideresnet50 & 0.78 & 0.015 & 0.22 & 0.418 & 0.544 & 0.865 \\ 
	  resnext101 & \textbf{0.80} & 0.019 & \textbf{0.20} & 0.369 & 0.505 & 0.881 \\ 
	  wideresnet101 & 0.78 & 0.021 & 0.22 & 0.336 & 0.468 & 0.874 \\ 
	  All Combined & \textbf{0.80} & \textbf{0.009} & \textbf{0.20} & 0.533 & \textbf{0.640} & \textbf{0.894} \\ 
	   \hline
	\end{tabular}
	}
	\caption{Average performance metrics by model architecture. Note that all metrics except ACC, which is multiclass accuracy, are for COVID-19 detection only (i.e., binary classification). TPR: true positive rate (or recall); FPR: false positive rate; FNR: false negative rate; PPV: positive predictive value (or precision); F1: F1-score; ACC: overall accuracy (TP+TN)/n.}
\label{tab:accmetrics}
\end{table}

Using the ensemble of models trained during this study, the FPR and FNR become 0.009 and 0.20, respectively. The ensemble predictions were carried out by simply summing the output from the last layer (softmax) of each network for a given image in the test hold-out set. The model ensemble offers no improvement over best performing models in FNR or FPR but improves the F1 score and multiclass accuracy to 0.640 (from 0.625) and 0.894 (from 0.884), respectively.

\section{Discussion}

The AI community is responding to the COVID-19 crisis and releasing publications at a rapid pace. Studies show promise in using AI in a clinical setting as a screening tool for diseases like COVID-19. While these results suggest that machine-learning techniques should only be used for COVID-19 screening (not diagnostic purposes), our intent with this study was to accelerate existing work for clinical augmentation purposes and provide insight into model selection for COVID-19 use cases. The use of AI in diagnostic procedures should be limited for use as a \emph{decision support} tool during screening, and diagnoses should \emph{not} rely on AI results alone.
 
Given the length of the incubation period and the variability in the symptom onset latency from infection, it is difficult to control for the time at which the image was acquired relative to the time of infection. However, efforts have been made to include an offset in COVID-19 imaging datasets by accounting for the number of days since the start of symptoms or hospitalization \cite{cohencovid192020}. Those who curate these datasets also must deal with inherent ambiguity in medical records such as image acquisition "after a few days" of symptoms (for example, Cohen et al. \cite{cohencovid192020} assume 5 days).

With the currently available datasets, AI engineers rely completely on clinical diagnoses and therefore assume that no false positive images exist; i.e., it is assumed that patients that tested positive for COVID-19 are indeed COVID-19 positive. Given an estimated false negative rate for COVID-19 tests are high (~10\%) \cite{yelin_evaluation_2020}, perhaps those testing positive can be safely assumed that they are indeed infected with the virus (currently, false positive RT-qPCR tests are not reported).

Deep learning architectures like those in this study have been translated for use in upper body CT images~\cite{gozes2020coronavirus}, and current studies demonstrate more reliable predictions than those from chest x-rays~\cite{li_artificial_2020,kassani2020automatic}. Reasons for this performance relate to the image detail that can be obtained from a chest CT as well as the size of the dataset used by Li et al. \cite{li_artificial_2020}, which contained images from 1296 COVID-19 patients allowing for a larger hold-out test size than what is currently available in x-ray (127 images at the time of this study) \cite{li_artificial_2020}. While CT scans provide enhanced screens, they are less accessible, more expensive, and less efficient than the chest x-ray due to the preparation time and scan time required. Consequently, CT falls short as a COVID-19 screening tool due to limitations including the complex mechanics and calibrations required for 3D geometrical renderings. Furthermore, only 25 CT scanners exist in the United States per million population (approx.) \cite{Castillo583}, and COVID-19 screens place an increased demand on top of existing demand for CT scans. Not only that, but an increased infection risk is an unfortunate corollary for non-COVID-19 patients requiring CT scans \cite{shi2020review}, especially due to the amount of CT essential equipment that interfaces with the patient compared to that for chest x-ray: not to mention the increased exposure for clinicians and technicians. If COVID-19 screens are needed \emph{en masse}, the chest x-ray is a promising, low-cost service that requires no moving parts and could be modified to meet a spike in demand.

While this study suffers from several limitations, accuracy metrics on a few tested architectures achieved results either consistent with or better than the current state-of-the-art in both x-ray and CT studies \cite{wangcovidnet2020, xu2020deep, li_artificial_2020, shi2020large, chen2020deep, jin2020development}. The results of this study, though limited to specific neural network architectures, suggest that transfer learning provides an efficient means to achieve high accuracy in detecting COVID-19. However, models remain highly dependent upon model architecture and vary depending on initial conditions and data augmentation steps. To better quantify the reliability of AI predictions in this context, our next goal is to implement segmentation techniques~\cite{qiu2020miniseg} and carry out a cross-validation protocol for each architecture and use a richer dataset. 

Model accuracy metrics indicate that more advancements are necessary before using AI for COVID-19 screening via x-ray. In our opinion, clinicians should not rely on solutions derived from architectures that have high (and high variability) FNR. Furthermore, inconsistent behavior in predictions between models with identical architectures injects doubt into its efficacy, which is unlikely to resolve until dataset limitations are worked out. Without a more abundant dataset, we do not expect deep learning approaches for COVID-19 screening to gain the reliability needed for clinical implementation. 

Finally, our aim is to encourage a focus on advancing the quality of x-ray screens for COVID-19 due to its efficiency over other means and to accelerate workflows that seek to leverage AI. We have made pre-trained model weights and the code for training freely available to the community through our github repository under a Creative Commons Attribution-NonCommercial 4.0 License (\href{https://covidresearch.ai/datasets/dataset?id=2}{covidresearch.ai/datasets}; \href{https://github.com/synthetaic/COVID19-IntraModel-Variability.git}{github.com/synthetaic}). We expect these model weights to provide significant improvements in model training efficiency as these public datasets continue to grow and evolve. We believe that AI results have the potential to achieve a degree of reliability that alleviates skepticism within the medical community regarding the use of chest x-ray and computer vision to screen COVID-19.

%\begin{thebibliography}{1}
%
%	\bibitem{kour2014real}
%	George Kour and Raid Saabne.
%	\newblock Real-time segmentation of on-line handwritten arabic script.
%	\newblock In {\em Frontiers in Handwriting Recognition (ICFHR), 2014 14th
%			International Conference on}, pages 417--422. IEEE, 2014.
%
%	\bibitem{kour2014fast}
%	George Kour and Raid Saabne.
%	\newblock Fast classification of handwritten on-line arabic characters.
%	\newblock In {\em Soft Computing and Pattern Recognition (SoCPaR), 2014 6th
%			International Conference of}, pages 312--318. IEEE, 2014.
%
%	\bibitem{hadash2018estimate}
%	Guy Hadash, Einat Kermany, Boaz Carmeli, Ofer Lavi, George Kour, and Alon
%	Jacovi.
%	\newblock Estimate and replace: A novel approach to integrating deep neural
%	networks with existing applications.
%	\newblock {\em arXiv preprint arXiv:1804.09028}, 2018.
%\end{thebibliography}
\section{Future Directions}

Models trained for this study had the task of detecting COVID-19 versus community acquired pneumonia and non-COVID-19 (healthy cases), which limits feature space to which the models are exposed. Future work should include many more image classifications to enable the network to learn features specific to a given pathology, which could provide the means to elucidate differentiable features of COVID-19 in chest x-ray. We aim to solve the data limitation problem in future work through numerical methods and data collection. 

\newcommand{\BIBdecl}{\footnotesize\setlength{\itemsep}{0pt}}
\bstctlcite{bstctl:etal, bstctl:nodash, bstctl:simpurl}
\bibliographystyle{IEEEtran}
% \footnotesize
%\bibliography{ref.bib} % commented by brian

%\bibliographystyle{unsrt}
\bibliography{references}  %%% Remove comment to use the external .bib file (using bibtex).

\begin{thebibliography}{10}
\providecommand{\url}[1]{#1}
\csname url@samestyle\endcsname
\providecommand{\newblock}{\relax}
\providecommand{\bibinfo}[2]{#2}
\providecommand{\BIBentrySTDinterwordspacing}{\spaceskip=0pt\relax}
\providecommand{\BIBentryALTinterwordstretchfactor}{4}
\providecommand{\BIBentryALTinterwordspacing}{\spaceskip=\fontdimen2\font plus
\BIBentryALTinterwordstretchfactor\fontdimen3\font minus
  \fontdimen4\font\relax}
\providecommand{\BIBforeignlanguage}[2]{{%
\expandafter\ifx\csname l@#1\endcsname\relax
\typeout{** WARNING: IEEEtran.bst: No hyphenation pattern has been}%
\typeout{** loaded for the language `#1'. Using the pattern for}%
\typeout{** the default language instead.}%
\else
\language=\csname l@#1\endcsname
\fi
#2}}
\providecommand{\BIBdecl}{\relax}
\BIBdecl

\bibitem{trackingproject}
\BIBentryALTinterwordspacing
``The covid tracking project (covidtracking.com).'' [Online]. Available:
  \url{https://covidtracking.com/}
\BIBentrySTDinterwordspacing

\bibitem{ai2020correlation}
T.~Ai, Z.~Yang, H.~Hou, C.~Zhan, C.~Chen, W.~Lv, Q.~Tao, Z.~Sun, and L.~Xia,
  ``Correlation of chest ct and rt-pcr testing in coronavirus disease 2019
  (covid-19) in china: a report of 1014 cases,'' \emph{Radiology}, p. 200642,
  2020.

\bibitem{shi2020review}
F.~Shi, J.~Wang, J.~Shi, Z.~Wu, Q.~Wang, Z.~Tang, K.~He, Y.~Shi, and D.~Shen,
  ``Review of artificial intelligence techniques in imaging data acquisition,
  segmentation and diagnosis for covid-19,'' \emph{arXiv preprint
  arXiv:2004.02731}, 2020.

\bibitem{cohencovid192020}
\BIBentryALTinterwordspacing
J.~P. Cohen, P.~Morrison, and L.~Dao, ``\BIBforeignlanguage{en}{{COVID}-19
  {Image} {Data} {Collection}},''
  \emph{\BIBforeignlanguage{en}{arXiv:2003.11597 [cs, eess, q-bio]}}, Mar.
  2020, arXiv: 2003.11597. [Online]. Available:
  \url{http://arxiv.org/abs/2003.11597}
\BIBentrySTDinterwordspacing

\bibitem{wangcovidnet2020}
\BIBentryALTinterwordspacing
L.~Wang and A.~Wong, ``{COVID}-net: A tailored deep convolutional neural
  network design for detection of {COVID}-19 cases from chest radiography
  images.'' [Online]. Available: \url{http://arxiv.org/abs/2003.09871}
\BIBentrySTDinterwordspacing

\bibitem{yan2020covid19}
Q.~Yan, B.~Wang, D.~Gong, C.~Luo, W.~Zhao, J.~Shen, Q.~Shi, S.~Jin, L.~Zhang,
  and Z.~You, ``Covid-19 chest ct image segmentation -- a deep convolutional
  neural network solution,'' 2020.

\bibitem{huang_clinical_2020}
\BIBentryALTinterwordspacing
C.~Huang, Y.~Wang, X.~Li, L.~Ren, J.~Zhao, Y.~Hu, L.~Zhang, G.~Fan, J.~Xu,
  X.~Gu, Z.~Cheng, T.~Yu, J.~Xia, Y.~Wei, W.~Wu, X.~Xie, W.~Yin, H.~Li, M.~Liu,
  Y.~Xiao, H.~Gao, L.~Guo, J.~Xie, G.~Wang, R.~Jiang, Z.~Gao, Q.~Jin, J.~Wang,
  and B.~Cao, ``Clinical features of patients infected with 2019 novel
  coronavirus in wuhan, china,'' vol. 395, no. 10223, pp. 497--506. [Online].
  Available:
  \url{https://linkinghub.elsevier.com/retrieve/pii/S0140673620301835}
\BIBentrySTDinterwordspacing

\bibitem{ng2020imaging}
M.-Y. Ng, E.~Y. Lee, J.~Yang, F.~Yang, X.~Li, H.~Wang, M.~M.-s. Lui, C.~S.-Y.
  Lo, B.~Leung, P.-L. Khong \emph{et~al.}, ``Imaging profile of the covid-19
  infection: radiologic findings and literature review,'' \emph{Radiology:
  Cardiothoracic Imaging}, vol.~2, no.~1, p. e200034, 2020.

\bibitem{born2020pocovidnet}
J.~Born, G.~Brändle, M.~Cossio, M.~Disdier, J.~Goulet, J.~Roulin, and
  N.~Wiedemann, ``Pocovid-net: Automatic detection of covid-19 from a new lung
  ultrasound imaging dataset (pocus),'' 2020.

\bibitem{apostolopoulos2020extracting}
I.~D. Apostolopoulos, S.~Aznaouridis, and M.~Tzani, ``Extracting possibly
  representative covid-19 biomarkers from x-ray images with deep learning
  approach and image data related to pulmonary diseases,'' 2020.

\bibitem{canhelp}
\BIBentryALTinterwordspacing
W.~D. Heaven, ``A neural network can help spot covid-19 in chest x-rays,''
  2020. [Online]. Available:
  \url{https://www.technologyreview.com/2020/03/24/950356/coronavirus-neural-network-can-help-spot-covid-19-in-chest-x-ray-pneumonia/}
\BIBentrySTDinterwordspacing

\bibitem{aihelps}
\BIBentryALTinterwordspacing
B.~Dickson, ``How ai is helping in the fight against covid-19,'' 2020.
  [Online]. Available:
  \url{https://www.pcmag.com/news/how-ai-is-helping-in-the-fight-against-covid-19}
\BIBentrySTDinterwordspacing

\bibitem{aicouldspot}
\BIBentryALTinterwordspacing
L.~Dormehl, ``A.i. could help spot telltale signs of coronavirus in lung
  x-rays,'' 2020. [Online]. Available:
  \url{https://www.digitaltrends.com/cool-tech/using-ai-to-spot-coronavirus-lung-damage/}
\BIBentrySTDinterwordspacing

\bibitem{kanne2020essentials}
J.~P. Kanne, B.~P. Little, J.~H. Chung, B.~M. Elicker, and L.~H. Ketai,
  ``Essentials for radiologists on covid-19: an update—radiology scientific
  expert panel,'' 2020.

\bibitem{bianco_benchmark_2018}
\BIBentryALTinterwordspacing
S.~Bianco, R.~Cadene, L.~Celona, and P.~Napoletano, ``Benchmark analysis of
  representative deep neural network architectures,'' vol.~6, pp.
  64\,270--64\,277. [Online]. Available: \url{http://arxiv.org/abs/1810.00736}
\BIBentrySTDinterwordspacing

\bibitem{NEURIPS2019_9015}
A.~Paszke, S.~Gross, F.~Massa, A.~Lerer, J.~Bradbury, G.~Chanan, T.~Killeen,
  Z.~Lin, N.~Gimelshein, L.~Antiga, A.~Desmaison, A.~Kopf, E.~Yang, Z.~DeVito,
  M.~Raison, A.~Tejani, S.~Chilamkurthy, B.~Steiner, L.~Fang, J.~Bai, and
  S.~Chintala, ``Pytorch: An imperative style, high-performance deep learning
  library,'' in \emph{Advances in Neural Information Processing Systems 32},
  H.~Wallach, H.~Larochelle, A.~Beygelzimer, F.~dAlche Buc, E.~Fox, and
  R.~Garnett, Eds.\hskip 1em plus 0.5em minus 0.4em\relax Curran Associates,
  Inc., 2019, pp. 8024--8035.

\bibitem{he2016deep}
K.~He, X.~Zhang, S.~Ren, and J.~Sun, ``Deep residual learning for image
  recognition,'' in \emph{Proceedings of the IEEE conference on computer vision
  and pattern recognition}, 2016, pp. 770--778.

\bibitem{corrZagoruykoK16}
\BIBentryALTinterwordspacing
S.~Zagoruyko and N.~Komodakis, ``Wide residual networks,'' \emph{CoRR}, vol.
  abs/1605.07146, 2016. [Online]. Available:
  \url{http://arxiv.org/abs/1605.07146}
\BIBentrySTDinterwordspacing

\bibitem{xie2017aggregated}
S.~Xie, R.~Girshick, P.~Doll{\'a}r, Z.~Tu, and K.~He, ``Aggregated residual
  transformations for deep neural networks,'' in \emph{Proceedings of the IEEE
  conference on computer vision and pattern recognition}, 2017, pp. 1492--1500.

\bibitem{howard2017mobilenets}
A.~G. Howard, M.~Zhu, B.~Chen, D.~Kalenichenko, W.~Wang, T.~Weyand,
  M.~Andreetto, and H.~Adam, ``Mobilenets: Efficient convolutional neural
  networks for mobile vision applications,'' \emph{arXiv preprint
  arXiv:1704.04861}, 2017.

\bibitem{huang2017densely}
G.~Huang, Z.~Liu, L.~Van Der~Maaten, and K.~Q. Weinberger, ``Densely connected
  convolutional networks,'' in \emph{Proceedings of the IEEE conference on
  computer vision and pattern recognition}, 2017, pp. 4700--4708.

\bibitem{deng2009imagenet}
J.~Deng, W.~Dong, R.~Socher, L.-J. Li, K.~Li, and L.~Fei-Fei, ``Imagenet: A
  large-scale hierarchical image database,'' in \emph{2009 IEEE conference on
  computer vision and pattern recognition}.\hskip 1em plus 0.5em minus
  0.4em\relax Ieee, 2009, pp. 248--255.

\bibitem{dietterich1998}
\BIBentryALTinterwordspacing
T.~G. Dietterich, ``Approximate statistical tests for comparing supervised
  classification learning algorithms,'' vol.~10, no.~7, pp. 1895--1923.
  [Online]. Available:
  \url{http://www.mitpressjournals.org/doi/10.1162/089976698300017197}
\BIBentrySTDinterwordspacing

\bibitem{everitt1977}
B.~Everitt.\hskip 1em plus 0.5em minus 0.4em\relax London: Chapman and Hall,
  1977.

\bibitem{rcitation}
\BIBentryALTinterwordspacing
{R Core Team}, \emph{R: A Language and Environment for Statistical Computing},
  R Foundation for Statistical Computing, Vienna, Austria, 2020. [Online].
  Available: \url{https://www.R-project.org/}
\BIBentrySTDinterwordspacing

\bibitem{ggplot2}
\BIBentryALTinterwordspacing
H.~Wickham, \emph{ggplot2: Elegant Graphics for Data Analysis}.\hskip 1em plus
  0.5em minus 0.4em\relax Springer-Verlag New York, 2016. [Online]. Available:
  \url{https://ggplot2.tidyverse.org}
\BIBentrySTDinterwordspacing

\bibitem{yelin_evaluation_2020}
\BIBentryALTinterwordspacing
I.~Yelin, N.~Aharony, E.~Shaer-Tamar, A.~Argoetti, E.~Messer, D.~Berenbaum,
  E.~Shafran, A.~Kuzli, N.~Gandali, T.~Hashimshony, Y.~Mandel-Gutfreund,
  M.~Halberthal, Y.~Geffen, M.~Szwarcwort-Cohen, and R.~Kishony,
  ``\BIBforeignlanguage{en}{Evaluation of {COVID}-19 {RT}-{qPCR} test in
  multi-sample pools},'' Infectious Diseases (except HIV/AIDS), preprint, Mar.
  2020. [Online]. Available:
  \url{http://medrxiv.org/lookup/doi/10.1101/2020.03.26.20039438}
\BIBentrySTDinterwordspacing

\bibitem{gozes2020coronavirus}
O.~Gozes, M.~Frid-Adar, N.~Sagie, H.~Zhang, W.~Ji, and H.~Greenspan,
  ``Coronavirus detection and analysis on chest ct with deep learning,'' 2020.

\bibitem{li_artificial_2020}
\BIBentryALTinterwordspacing
L.~Li, L.~Qin, Z.~Xu, Y.~Yin, X.~Wang, B.~Kong, J.~Bai, Y.~Lu, Z.~Fang,
  Q.~Song, K.~Cao, D.~Liu, G.~Wang, Q.~Xu, X.~Fang, S.~Zhang, J.~Xia, and
  J.~Xia, ``Artificial intelligence distinguishes {COVID}-19 from community
  acquired pneumonia on chest {CT},'' p. 200905. [Online]. Available:
  \url{http://pubs.rsna.org/doi/10.1148/radiol.2020200905}
\BIBentrySTDinterwordspacing

\bibitem{kassani2020automatic}
S.~H. Kassani, P.~H. Kassasni, M.~J. Wesolowski, K.~A. Schneider, and
  R.~Deters, ``Automatic detection of coronavirus disease (covid-19) in x-ray
  and ct images: A machine learning-based approach,'' 2020.

\bibitem{Castillo583}
\BIBentryALTinterwordspacing
M.~Castillo, ``The industry of ct scanning,'' \emph{American Journal of
  Neuroradiology}, vol.~33, no.~4, pp. 583--585, 2012. [Online]. Available:
  \url{http://www.ajnr.org/content/33/4/583}
\BIBentrySTDinterwordspacing

\bibitem{xu2020deep}
X.~Xu, X.~Jiang, C.~Ma, P.~Du, X.~Li, S.~Lv, L.~Yu, Y.~Chen, J.~Su, G.~Lang
  \emph{et~al.}, ``Deep learning system to screen coronavirus disease 2019
  pneumonia,'' \emph{arXiv preprint arXiv:2002.09334}, 2020.

\bibitem{shi2020large}
F.~Shi, L.~Xia, F.~Shan, D.~Wu, Y.~Wei, H.~Yuan, H.~Jiang, Y.~Gao, H.~Sui, and
  D.~Shen, ``Large-scale screening of covid-19 from community acquired
  pneumonia using infection size-aware classification,'' \emph{arXiv preprint
  arXiv:2003.09860}, 2020.

\bibitem{chen2020deep}
J.~Chen, L.~Wu, J.~Zhang, L.~Zhang, D.~Gong, Y.~Zhao, S.~Hu, Y.~Wang, X.~Hu,
  B.~Zheng \emph{et~al.}, ``Deep learning-based model for detecting 2019 novel
  coronavirus pneumonia on high-resolution computed tomography: a prospective
  study,'' \emph{medRxiv}, 2020.

\bibitem{jin2020development}
C.~Jin, W.~Chen, Y.~Cao, Z.~Xu, X.~Zhang, L.~Deng, C.~Zheng, J.~Zhou, H.~Shi,
  and J.~Feng, ``Development and evaluation of an ai system for covid-19
  diagnosis,'' \emph{medRxiv}, 2020.

\bibitem{qiu2020miniseg}
Y.~Qiu, Y.~Liu, and J.~Xu, ``Miniseg: An extremely minimum network for
  efficient covid-19 segmentation,'' 2020.

\end{thebibliography}

\end{document}